\documentclass[10pt,journal,twocolumn]{IEEEtran}
\IEEEoverridecommandlockouts
\ifCLASSINFOpdf
\else
\fi
\usepackage{amsmath,amssymb,graphicx}
\usepackage{mathtools}
\usepackage{algpseudocode}
\usepackage{algorithm}
\usepackage{color}
\usepackage{bm}
\usepackage{amsfonts}
\usepackage{amscd}
\usepackage[mathcal]{eucal}
\usepackage{epsfig}
\usepackage{caption}
\usepackage{array}
\usepackage{eqparbox}
\usepackage{enumerate}
\usepackage{stackrel}
\usepackage{pdflscape}
\usepackage{lineno}
\usepackage{verbatim}
\usepackage{cite}
\usepackage{afterpage}

\newcommand{\ie}{{\em i.e.}}
\newcommand{\etc}{{\em etc}}

\newcommand{\iid}{i.i.d.}

\newcommand{\secref}[1]{Section~\ref{#1}}

\newcommand{\qed}{\nobreak \ifvmode \relax \else
      \ifdim\lastskip<1.5em \hskip-\lastskip
      \hskip1.5em plus0em minus0.5em \fi \nobreak
      \vrule height0.75em width0.5em depth0.25em\fi}

\begin{document}
\title{On the Relationship Between Measures of Relative Efficiency for Random Signal Detection}
\author{
\authorblockN{Nagananda Kyatsandra}
\authorblockA{LTCI, T\'el\'ecom ParisTech,\\ Institut Mines - T\'el\'ecom,\\
Paris 75013, France.\\
Email: \texttt{nkyatsandra@enst.fr}}
}

\author{K. G. Nagananda
\thanks{K. G. Nagananda was with LTCI, T\'el\'ecom ParisTech, Institut Mines - T\'el\'ecom, Paris 75013, France,  E-mail: \texttt{nkyatsandra@enst.fr}. The work was supported by European Research Council under grant agreement 715111.}
} 

\maketitle
\thispagestyle{empty}
\pagestyle{empty}

\begin{abstract}
Relative efficiency (RE), the Pitman asymptotic relative efficiency (ARE) and efficacy are important relative performance measures of signal detection techniques. These measures allow comparing two detectors in terms of the relative sample sizes they require to achieve the same prescribed level of false alarm and detection probabilities. While the finite-sample-size measure RE is useful to analyze the small sample behavior of detectors, in practice it is difficult to compute. In the limit as the signal strength approaches zero at an appropriate rate, the RE converges to an asymptotic limit only for very large sample sizes. This limiting ratio of the number of samples is the ARE, which lends analytical tractability, but does not provide insights into the finite sample behavior of detectors which is important for practical applications. This led researchers to study the convergence of RE to ARE, and has been well-reported for the problem of constant signal detection in additive, independent, and identically distributed noise. When the signal to be detected is random ({\ie}, unknown), such a convergence analysis is lacking in the literature and is the focus of this paper. A relationship between RE and ARE for random signal detection is established. We use the higher-order terms in the Taylor series expansion of the mean of the test statistic under the alternative hypothesis to derive this formula. We present preliminary remarks on the convergence of RE to ARE for random signals in comparison to that for constant signal detection.
\end{abstract}

\begin{IEEEkeywords}
Relative efficiency, asymptotic relative efficiency, random signal detection.
\end{IEEEkeywords}
\IEEEpeerreviewmaketitle

\vspace{-0.1in}
\section{Introduction}\label{sec:introduction}
The ratio of the number of samples needed by two detectors to achieve the same prescribed level of detection performance is referred to as the relative efficiency (RE) of the two detectors. The limit of the RE as the sample sizes approach infinity and the signal strengths approach zero is referred to as the Pitman asymptotic relative efficiency (ARE) \cite{Eeden1963}, \cite{Poor1988}. The ARE is a widely employed comparative measure of the performance of two detectors in optimal detector quantization \cite{Kassam1985}, robust detection \cite{Kassam1985a}, nonparametric detection \cite{Thomas1970}, rank tests \cite{Papantoni-Kazakos1975}, score tests \cite{Tosteson1988}, {\etc}. The ARE is a large-sample measure which offers the advantage of analytical tractability, since it enables the use of the central limit theorem \cite{Vaart1998,Bening2000,Nikitin2009}. 

In practice, only finite number of data samples are available, which raises concerns about the validity of ARE as an appropriate measure of the efficiency of the detector. In such cases, the RE (a finite-sample-size measure) seems a reasonable performance index. However, it is well-known that computing the RE suffers from mathematical difficulties \cite{Bening2000}, thus motivating alternative viewpoints. One insightful approach is the analysis of the convergence behavior of RE to ARE. Essentially, it has been shown that RE has a very slow rate of convergence to ARE, and convergence may be from above or from below \cite{Miller1975,Marks1978,Michalsky1982,Liu1983,Dadi1987,Helstrom1989}. It has also been shown that the convergence behavior can be predicted systematically for a class of detectors by deriving estimates of RE \cite{Blum1991}. Asymptotic expansions were used in the central limit theorem to obtain more accurate indices of RE for some practically important detectors in \cite{Burnashev1995}. RE versus ARE analysis for sequential detectors has also been reported \cite{Bechhofer1960,Sakaguchi1967,Tantaratana1978}. 

The aforementioned papers considered detection of constant signals in additive, independent, and identically distributed noise ({\iid}). To the best of our knowledge, the study of the relationship between relative performance measures for the detection of random signals in noise does not exist in the literature. With random signal detection witnessing enormous developments in the recent past with widespread applications (for example,\cite{Wimalajeewa2018,Gezici2018,Mian2019,Tugnait2019,Lau2019}), we are motivated to consider the detection of random signals in {\iid} Gaussian noise, and develop a formula that relates RE and ARE for this case. The formula is based on the higher-order terms in the Taylor series expansion of the mean of the test statistic under the alternative hypothesis. We provide comparisons between our formula and the one developed in \cite{Blum1991}, which was for constant signal detection. At first glance, our formula looks identical to the one in \cite{Blum1991}, however, upon closer inspection we see that the random nature of the signal makes it difficult to predict the convergence behavior of RE to ARE. Using this formula, we provide preliminary comparisons between the convergence of RE to ARE for constant and random signal detections. 

In \secref{sec:preliminaries}, we introduce the system model and provide formulas for RE, efficacy and ARE. In \secref{sec:np_random}, we characterize the performance of the random signal detection by deriving expressions for the probabilities of false alarm and detection. The formula that relates RE to ARE is derived in \secref{sec:re_are}. Some comparisons between the convergence of RE to ARE for constant and random signal detection problems are provided in \secref{sec:remarks}. 

\vspace{-0.1in}
\section{Preliminaries}\label{sec:preliminaries}
The detection problem considered in this paper is to resolve $H_0: x[n] = w[n]$ and $H_1: x[n] = s + w[n]$, $n = 0,\dots,N-1$. We assume the signal $S \sim \mathcal(\mu_1, \sigma_1^2)$ and the noise $W \sim \mathcal(\mu_0, \sigma_0^2)$. Therefore, under $H_0$, $X \sim \mathcal{N}(\mu_0, \sigma_0^2)$ and under $H_1$, $X \sim \mathcal{N}(\mu_0 + \mu_1, \sigma_0^2 + \sigma_1^2)$.

Given a hypothesis test, we denote the probabilities of false alarm and detection by $P_F$ and $P_D$, respectively. Let $T_A(\bm{x})$ and $T_B(\bm{x})$ be the test statistics of two different tests with sample sizes $N_A$ and $N_B$, respectively, required to attain the prescribed $(P_F, P_D)$ to distinguish between any two hypothesis $H_0$ and $H_1$. The RE of $T_B(\bm{x})$ with respect to $T_A(\bm{x})$ is given by $\text{RE}_{A, B} = \frac{N_B}{N_A}$, and and the ARE is given by $\text{ARE}_{A, B} = \lim_{N \rightarrow \infty}\frac{N_B}{N_A}$. The efficacy is used to measure the resolution capability of a detector, and is given by \cite{Song1990}
\begin{eqnarray}
\sqrt{\xi} \!\! &=& \!\!\!\! \lim\limits_{N \rightarrow \infty} \frac{\frac{d^{\nu} \mathbb{E}[T_N(\bm{x})|H_1]}{d s^{\nu}}\big|_{s = 0}}{\sqrt{N}\sqrt{\text{var}\left[T_N(\bm{x})|H_0\right]}},
\label{eq:efficacy}
\end{eqnarray}
$\mathbb{E}[\cdot]$ and $\text{var}(\cdot)$ denote mean and variance, respectively, and $\nu$ is the smallest order for which the derivative at $s = 0$ is nonzero. It is commonly assumed that there exists at least one $\nu \neq 0$ for which the derivative is nonzero \cite[Theorem 13.2.1]{Lehmann2005}. The ARE is generally expressed as the ratio of efficacies:
\begin{eqnarray}
\text{ARE}_{A, B} = \frac{\lim\limits_{N \rightarrow \infty} \frac{\frac{d^{\nu} \mathbb{E}[T_{N_A}(\bm{x})|H_1]}{d s^{\nu}}\big|_{s = 0}}{\sqrt{N_A}\sqrt{ \text{var}\left[T_{N_A}(\bm{x})|H_0\right]}}}{\lim\limits_{N \rightarrow \infty} \frac{\frac{d^{\nu} \mathbb{E}[T_{N_B}(\bm{x})|H_1]}{d s^{\nu}}\big|_{s = 0}}{\sqrt{N_B}\sqrt{ \text{var}\left[T_{N_B}(\bm{x})|H_0\right]}}}.
\end{eqnarray}

\section{Hypothesis test for random signal detection}\label{sec:np_random}
In this section, we characterize the performance of the Neyman-Pearson (NP) test to resolve $H_0$ from $H_1$. The standard practice for random signal detection is to assume $\mu_0 = \mu_1 = 0$; see \cite{Poor1988}. In the context of this paper, this assumption does not allow a complete characterization of the performance of the detector, required to analyze how our result compares with the convergence of RE to ARE analysis presented in \cite{Blum1991}. However, in the next section, we let $\mu_0 = 0$. So, we first provide the $(P_D, P_F)$ characterization by letting both $\mu_0$ and $\mu_1$ to be nonzero. We fix the probability of false alarm $P_F = \alpha$. The log-likelihood ratio test is given by
\begin{eqnarray}
\sum\limits_{n=0}^{N-1}x^2[n] + \frac{2(\sigma^2_1\mu_0 + \sigma^2_0\mu_1)}{\sigma^2_1}\sum\limits_{n=0}^{N-1}x[n]  \stackrel[H_0]{H_1}{\gtrless} \frac{\gamma'}{\sigma^2_1},
\label{eq:test}
\end{eqnarray}
where $\gamma' = 2(\sigma^2_1+\sigma^2_0)\sigma^2_0  \left[\ln(\gamma) - \frac{N}{2}\ln \left(\frac{\sigma^2_0}{\sigma^2_1+\sigma^2_0}\right) \right] + \sigma^2_0\mu_1^2 - \sigma_1^2\mu_0^2 + 2\sigma_0^2\mu_0\mu_1$. The value of $\gamma'$ (and hence $\gamma$) can easily be obtained for a fixed probability of false alarm using the Neyman-Pearson lemma as will shown in the sequel. Now letting $\delta = \frac{2(\sigma^2_1\mu_0 + \sigma^2_0\mu_1)}{\sigma^2_1}$, the test simplifies to 
\begin{eqnarray}
T(\bm{x}) = \sum\limits_{n=0}^{N-1}x^2[n] + \delta\sum\limits_{n=0}^{N-1}x[n]  \stackrel[H_0]{H_1}{\gtrless} \frac{\gamma'}{\sigma^2_1}.
\label{eq:test1}
\end{eqnarray}

To characterize the performance of the hypothesis test, we make a slight modification to the test statistic $T(\bm{x})$. Under hypothesis $H_0$, we consider the statistic $\frac{T(\bm{x})}{\sigma_0^2}$:
\begin{eqnarray}
\frac{T(\bm{x})}{\sigma^2_0} = \frac{\sum\limits_{n=0}^{N-1}x^2[n]}{\sigma^2_0} + \frac{\delta\sum\limits_{n=0}^{N-1}x[n]}{\sigma^2_0}.
\end{eqnarray}
The distribution of $\frac{T(\bm{x})}{\sigma^2}$ is the sum of the distributions of $\frac{\sum\limits_{n=0}^{N-1}x^2[n]}{\sigma^2_0}$ and $ \frac{\delta\sum\limits_{n=0}^{N-1}x[n]}{\sigma^2_0}$. It can be seen that 
\begin{eqnarray}
\frac{\sum\limits_{n=0}^{N-1}x^2[n]}{\sigma^2_0} \sim \chi^2_N (N\mu_0^2),
\end{eqnarray}
where $\chi^2_N (N\mu_0^2)$ denotes the non-central Chi-squared distribution with $N$ degrees of freedom and the non-centrality parameter $\lambda_0 = \sum\limits_{i=1}^{N}\mu_{0}^2 = N\mu_0^2$, while 
\begin{eqnarray}
\frac{\delta\sum\limits_{n=0}^{N-1}x[n]}{\sigma^2_0} \sim \mathcal{N}\left(\frac{N\mu_0\delta}{\sigma_0^2},  \frac{N\delta^2}{\sigma_0^2}\right), 
\end{eqnarray}
since we have 
\begin{eqnarray}
\mathbb{E}_{H_0}\left[\frac{\delta\sum\limits_{n=0}^{N-1}x[n]}{\sigma^2_0}\right] = \frac{\delta}{\sigma_0^2}\mathbb{E}\left[\sum\limits_{n=0}^{N-1}x[n] \right] = \frac{N\mu_0\delta}{\sigma_0^2}, \\
\text{var}_{H_0}\left[\frac{\delta\sum\limits_{n=0}^{N-1}x[n]}{\sigma^2_0}\right] = \frac{\delta^2}{\sigma_0^4} \text{var}\left[\sum\limits_{n=0}^{N-1}x[n]\right] = \frac{N\delta^2}{\sigma_0^2}.
\end{eqnarray}

Under hypothesis $H_1$, we consider $\frac{T(\bm{x})}{\sigma_0^2 + \sigma_1^2}$:
\begin{eqnarray}
\frac{T(\bm{x})}{\sigma_0^2 + \sigma_1^2} = \frac{\sum\limits_{n=0}^{N-1}x^2[n]}{\sigma_0^2 + \sigma_1^2} + \frac{\delta\sum\limits_{n=0}^{N-1}x[n]}{\sigma_0^2 + \sigma_1^2}.
\end{eqnarray}
The distribution of $\frac{T(\bm{x})}{\sigma_0^2 + \sigma_1^2}$ is the sum of the distributions of $\frac{\sum\limits_{n=0}^{N-1}x^2[n]}{\sigma_0^2 + \sigma_1^2}$ and $\frac{\delta\sum\limits_{n=0}^{N-1}x[n]}{\sigma_0^2 + \sigma_1^2}$, with
\begin{eqnarray}
\frac{\sum\limits_{n=0}^{N-1}x^2[n]}{\sigma_0^2 + \sigma_1^2} \sim \chi^2_N \left(N(\mu_0 + \mu_1)^2\right),
\end{eqnarray}
$\chi^2_N \left(N(\mu_0 + \mu_1)^2\right)$ denotes the non-central Chi-squared distribution with $N$ degrees of freedom and the non-centrality parameter $\lambda_1 = \sum\limits_{i=1}^{N}(\mu_{0}+\mu_1)^2 = N(\mu_0 + \mu_1)^2$, and
\begin{eqnarray}
\frac{\delta\sum\limits_{n=0}^{N-1}x[n]}{\sigma_0^2 + \sigma_1^2} \sim \mathcal{N}\left(\frac{N(\mu_0+\mu_1)\delta}{\sigma_0^2 + \sigma_1^2}, \frac{N\delta^2}{\sigma_0^2 + \sigma_1^2}\right), 
\end{eqnarray}
since 
\begin{eqnarray}
\mathbb{E}_{H_1}\left[\frac{\delta\sum\limits_{n=0}^{N-1}x[n]}{\sigma_0^2 + \sigma_1^2}\right] \!\!\!\!\!\! &=& \!\!\!\!\!\! \frac{N(\mu_0+\mu_1)\delta}{\sigma_0^2 + \sigma_1^2}, \\
\text{var}_{H_1}\left[\frac{\delta\sum\limits_{n=0}^{N-1}x[n]}{\sigma_0^2 + \sigma_1^2}\right] \!\!\!\!\!\! &=& \!\!\!\!\!\! \frac{N\delta^2}{\sigma_0^2 + \sigma_1^2}.
\end{eqnarray}

In summary, 
\begin{eqnarray}
\frac{T(\bm{x})}{\sigma^2_0}\bigg|H_0 \!\!\!\!\! &\sim& \!\!\!\!\! \chi^2_N (N\mu_0^2) + \mathcal{N}\left(\frac{N\mu_0\delta}{\sigma_0^2},  \frac{N\delta^2}{\sigma_0^2}\right), \\
\nonumber \frac{T(\bm{x})}{\sigma_0^2 + \sigma_1^2}\bigg|H_1 \!\!\!\!\! &\sim& \!\!\!\!\! \chi^2_N \left(N(\mu_0 + \mu_1)^2\right) \\ &&  + \mathcal{N}\left(\frac{N(\mu_0+\mu_1)\delta}{\sigma_0^2 + \sigma_1^2}, \frac{N\delta^2}{\sigma_0^2 + \sigma_1^2}\right).
\end{eqnarray}

In the study of RE versus ARE, it is generally assumed that the signal strength $s \rightarrow 0$ and the number of samples $N \rightarrow \infty$. For the random signal detection problem considered in this paper, the first assumption is automatically relaxed, while the assumption of very large sample sizes is retained. For $N \rightarrow \infty$, we have 
\begin{eqnarray}
\chi^2_N (N\mu_0^2) \!\!\!\!\! &\sim& \!\!\!\!\! \mathcal{N}\left(N+ N\mu_0^2,  2(N + 2N\mu_0^2)\right), \\
\nonumber \chi^2_N \left(N(\mu_0 + \mu_1)^2\right) \!\!\!\!\! &\sim& \!\!\!\!\! \mathcal{N}\left(N+ N(\mu_0 + \mu_1)^2,\right. \\ && \left. 2(N + 2N(\mu_0 + \mu_1)^2)\right).
\end{eqnarray}

After simplification, the null and the alternative distributions are expressed as below: 
\begin{align*}
\begin{cases}
H_0:  \frac{\frac{T(\bm{x})}{\sigma^2_0} - \left(N+ N\mu_0^2 + \frac{N\mu_0\delta}{\sigma_0^2}\right)}{\sqrt{2(N + 2N\mu_0^2) + \frac{N\delta^2}{\sigma_0^2}}} \sim  \mathcal{N}\left(0, 1\right), \\
H_1: \frac{\frac{T(\bm{x})}{(\sigma_0^2 + \sigma_1^2)} - \left(N+ N(\mu_0 + \mu_1)^2 + \frac{N(\mu_0+\mu_1)\delta}{\sigma_0^2 + \sigma_1^2} \right)}{\sqrt{2(N + 2N(\mu_0 + \mu_1)^2) + \frac{N\delta^2}{\sigma_0^2 + \sigma_1^2}}}\sim \mathcal{N}\left(0, 1\right),
\end{cases}
\end{align*}
which are used to derive expressions for $P_F$ and $P_D$.

For a fixed $P_F = \alpha$, using the Neyman-Pearson lemma, the threshold $\gamma'$ is given by
\begin{eqnarray}
\nonumber P_F \!\!\!\!\! &=& \!\!\!\!\! p\left(\frac{T(\bm{x})}{\sigma_0^2} > \frac{\gamma'}{\sigma^2_0\sigma_1^2} \bigg|H_0 \right) \\
\nonumber \!\!\!\!\! &=& \!\!\!\!\! Q\left(\frac{\frac{\gamma'}{\sigma^2_0\sigma^2_1} - \left(N+ N\mu_0^2 + \frac{N\mu_0\delta}{\sigma_0^2}\right)}{\sqrt{2(N + 2N\mu_0^2) + \frac{N\delta^2}{\sigma_0^2}}} \right). \\
\nonumber \gamma' \!\!\!\!\! &=& \!\!\!\!\! \sigma^2_0\sigma^2_1\left[Q^{-1}(\alpha)\sqrt{2(N + 2N\mu_0^2)  + \frac{N\delta^2}{\sigma_0^2}} + \right. \\ && \left. \left(N+ N\mu_0^2 + \frac{N\mu_0\delta}{\sigma_0^2}\right)\right].
\end{eqnarray}
The threshold $\gamma'$ is then used to obtain the probability of detection $P_D$: 
\begin{align}
\nonumber P_D = p\left(\frac{T(\bm{x})}{\sigma_0^2 + \sigma_1^2} > \frac{\gamma'}{\sigma^2_1(\sigma_0^2 + \sigma_1^2)} \bigg|H_1 \right) \\
=  Q\left(\frac{\frac{\gamma'}{\sigma^2_1(\sigma_0^2 + \sigma_1^2)} - \left(N+ N(\mu_0 + \mu_1)^2 + \frac{N(\mu_0+\mu_1)\delta}{\sigma_0^2 + \sigma_1^2} \right)}{\sqrt{2(N + 2N(\mu_0 + \mu_1)^2) + \frac{N\delta^2}{\sigma_0^2 + \sigma_1^2}}} \right). 
\end{align}

\section{Relationship between RE and ARE}\label{sec:re_are}
We now derive a formula which relates RE to ARE for random signal detection. To derive this formula, we need to express $P_D$ in a form suitable for making use of the definitions of RE, efficacy and ARE. As an aid to this expression, we let $\mu_0 = 0$. We could have simply let $\mu_0 = 0$ in the previous section, however, that does not lead to a systematic development of the formula. With $\mu_0 = 0$, we get $\delta = \frac{2\sigma^2_0\mu_1}{\sigma^2_1}$ and 
\begin{eqnarray}
\begin{cases}
H_0: \frac{\frac{T(\bm{x})}{\sigma^2_0} - N}{\sqrt{2N + \frac{N\delta^2}{\sigma_0^2}}} \sim \mathcal{N}\left(0, 1\right), \\
H_1: \frac{\frac{T(\bm{x})}{(\sigma_0^2 + \sigma_1^2)} - \left(N+ N\mu_1^2 + \frac{N\mu_1\delta}{\sigma_0^2 + \sigma_1^2} \right)}{\sqrt{2\left(N + 2N\mu_1^2\right) + \frac{N\delta^2}{\sigma_0^2 + \sigma_1^2}}}\sim \mathcal{N}\left(0, 1\right),
\end{cases}
\end{eqnarray}
yielding simplified expressions for the threshold $\gamma'$ and the probability of detection $P_D$:
\begin{eqnarray}
\gamma' \!\!\!\!\! &=& \!\!\!\!\! \sigma^2_0\sigma^2_1\left[Q^{-1}(\alpha)\sqrt{2N + \frac{N\delta^2}{\sigma_0^2}} + N \right] \\
\nonumber P_D \!\!\!\!\! &=& \!\!\!\!\! Q\left(\frac{\frac{\sigma^2_0\sigma^2_1\left[Q^{-1}(\alpha)\sqrt{2N + \frac{N\delta^2}{\sigma_0^2}} + N \right]}{\sigma^2_1(\sigma_0^2 + \sigma_1^2)}}{\sqrt{2(N + 2N\mu_1^2) + \frac{N\delta^2}{\sigma_0^2 + \sigma_1^2}}}  \right. \\ \nonumber && \left. - \frac{\left(N+ N\mu_1^2 + \frac{N\mu_1\delta}{\sigma_0^2 + \sigma_1^2} \right)}{\sqrt{2(N + 2N\mu_1^2) + \frac{N\delta^2}{\sigma_0^2 + \sigma_1^2}}} \right) \\
\nonumber &=& \!\!\!\!\! 1 - \Phi\left(\frac{\sigma^2_0\sqrt{2N + \frac{N\delta^2}{\sigma_0^2}} \Phi^{-1}(1 - \alpha)}{(\sigma_0^2 + \sigma_1^2)\sqrt{2(N + 2N\mu_1^2) + \frac{N\delta^2}{(\sigma_0^2 + \sigma_1^2)}}}  - \right. \\  && \left. \frac{(\sigma_0^2 + \sigma_1^2)\left[N+ N\mu_1^2 + \frac{N\mu_1\delta}{(\sigma_0^2 + \sigma_1^2)} \right] - N\sigma_0^2}{(\sigma_0^2 + \sigma_1^2)\sqrt{2(N + 2N\mu_1^2) + \frac{N\delta^2}{(\sigma_0^2 + \sigma_1^2)}}}\right),
\label{eq:PD1}
\end{eqnarray}
where $\Phi(\cdot)$ is the cumulative distribution function of the standard normal distribution. In general, we can write $H_0: T(\bm{x}) \sim \mathcal{N}\left(T_{\mu_{0, N}}, T_{\sigma^2_{0, N}}\right)$ and $H_1: T(\bm{x}) \sim \mathcal{N}\left(T_{\mu_{1, N}}, T_{\sigma^2_{1, N}}\right)$, where $T_{\mu_{0, N}}$ (resp. $T_{\mu_{1, N}}$) and $T_{\sigma^2_{0, N}}$ (resp. $T_{\sigma^2_{1, N}}$) denote the mean and variance of the test statistic $T(\bm{x})$ with sample size $N$ under hypothesis $H_0$ (resp. $H_1$). For the random signal detection problem considered here, we have 
\begin{eqnarray}
T_{\mu_{0, N}} \!\!\!\!\! &=& \!\!\!\!\! N\sigma^2_0, \\
T_{\mu_{1, N}} \!\!\!\!\! &=& \!\!\!\!\! (\sigma_0^2 + \sigma_1^2)\left[N+ N\mu_1^2 + \frac{N\mu_1\delta}{\sigma_0^2 + \sigma_1^2} \right], \\
T_{\sigma^2_{0, N}} \!\!\!\!\! &=& \!\!\!\!\! \sigma^4_0\left[2N + \frac{N\delta^2}{\sigma_0^2}\right], \\
T_{\sigma^2_{1, N}} \!\!\!\!\! &=& \!\!\!\!\! (\sigma_0^2 + \sigma_1^2)^2\left[2\left(N + 2N\mu_1^2\right) + \frac{N\delta^2}{\sigma_0^2 + \sigma_1^2}\right].
\end{eqnarray}
For notational convenience, we write $\sqrt{T_{\sigma^2_{0, N}}} = T_{\sigma_{0, N}}$ and $\sqrt{T_{\sigma^2_{1, N}}} = T_{\sigma_{1, N}}$ to denote the standard deviation of the test statistic under the null and alternative hypotheses, respectively. Therefore, \eqref{eq:PD1} can be written in generic form as follows:
\begin{eqnarray}
P_D \!\!\!\!\! &=& \!\!\!\!\! 1 - \Phi\left(\frac{T_{\sigma_{0, N}}}{T_{\sigma_{1, N}}}\Phi^{-1}(1 - \alpha) - \frac{T_{\mu_{1, N}} - T_{\mu_{0, N}}}{T_{\sigma_{1, N}}}\right).
\label{eq:relation}
\end{eqnarray}

Equation \eqref{eq:relation} is central to establish a relationship between RE and ARE, which will further be used to analyze the convergence of RE to ARE. Consider two detectors $A$ and $B$ with sample sizes $N_A$ and $N_B$, respectively, to achieve the same probability of detection for a given false alarm probability $\alpha$. The arguments of the cumulative distribution function $\Phi$ are equated to obtain 
\begin{align}
\nonumber \frac{T_{\sigma_{0, N_A}}}{T_{\sigma_{1, N_A}}}\Phi^{-1}(1 - \alpha) - \frac{T_{\mu_{1, N_A}} - T_{\mu_{0, N_A}}}{T_{\sigma_{1, N_A}}} = \\ \frac{T_{\sigma_{0, N_B}}}{T_{\sigma_{1, N_B}}}\Phi^{-1}(1 - \alpha) - \frac{T_{\mu_{1, N_B}} - T_{\mu_{0, N_B}}}{T_{\sigma_{1, N_B}}}.
\label{eq:relation1}
\end{align}

To relate RE and ARE, we consider the Taylor series expansion of the mean of each detector around the mean of the random signal $S$, {\ie}, around the point $\mu_1$:
\begin{eqnarray*}
T_{\mu_{1, N}}  \!\!\!\!\! &=& \!\!\!\!\!  T_{\mu_{1, N}}(\mu_1) + \sum_{k=1}^{\infty}\frac{1}{k\,!}(s - \mu_1)^k\frac{d^k T_{\mu_{1, N}}}{ds^k}\bigg|_{s = \mu_1},
\end{eqnarray*}
where $N = N_A$ for detector $A$ and $N = N_B$ for detector $B$. 

The efficacy is essentially is a measure of the resolution capability of the detector, {\ie}, we are interested in the situation where $\mu_1 \rightarrow \mu_0$. Note that, we have assumed $\mu_0 = 0$; therefore, we are basically interested in the efficacy of the detector for the case where $\mu_1 \rightarrow 0$ leading to 
\begin{eqnarray}
T_{\mu_{1, N}}  \!\!\!\!\! &=& \!\!\!\!\!  T_{\mu_{1, N}}(0) + \sum_{k=1}^{\infty}\frac{1}{k\,!}(s)^k\frac{d^k T_{\mu_{1, N}}}{ds^k}\bigg|_{s = 0},
\end{eqnarray}

The efficacy of detector $A$ is given by
\begin{eqnarray}
\sqrt{\xi_A} = \lim\limits_{N_A \rightarrow \infty} \frac{T^{(\nu)}_{\mu_{0, N_A}}}{\sqrt{N_A}T_{\sigma_{0, N_A}}},
\end{eqnarray}
$\nu$ is the smallest order for which the derivative at $s = 0$ is nonzero. Similarly for $\sqrt{\xi_B}$. In the following, we denote by $\Delta_{H_1, N}(s)$ accounting for higher order derivatives not equal to zero, where $N = N_A$ or $N_B$. First, we express \eqref{eq:relation} using terms of the Taylor series expansion as follows:
\begin{align*} 
\Phi^{-1}(1 - \alpha)\left[\frac{T_{\sigma_{0, N_A}}}{T_{\sigma_{1, N_A}}} -  \frac{T_{\sigma_{0, N_B}}}{T_{\sigma_{1, N_B}}}\right]  =  \frac{T_{\mu_{1, N_A}} - T_{\mu_{0, N_A}}}{T_{\sigma_{1, N_A}}} \\ - \frac{T_{\mu_{1, N_B}} - T_{\mu_{0, N_B}}}{T_{\sigma_{1, N_B}}}  \\
=  \frac{T_{\mu_{0, N_A}} + \frac{1}{\nu\,!}s^{\nu} T^{(\nu)}_{\mu_{0, N_A}} + \Delta_{H_1, N_A}(s) - T_{\mu_{0, N_A}}}{T_{\sigma_{1, N_A}}} \\  - \frac{T_{\mu_{0, N_B}} + \frac{1}{\nu\,!}s^{\nu} T^{(\nu)}_{\mu_{0, N_B}} + \Delta_{H_1, N_B}(s) - T_{\mu_{0, N_B}}}{T_{\sigma_{1, N_B}}} \\
= \frac{1}{\nu\,!}s^{\nu} \sqrt{\xi_A} \frac{T_{\sigma_{0, N_A}}}{T_{\sigma_{1, N_A}}} \sqrt{N_A} + \frac{\Delta_{H_1, N_A}(s)}{T_{\sigma_{1, N_A}}} \\ - \frac{1}{\nu\,!}s^{\nu} \sqrt{\xi_B} \frac{T_{\sigma_{0, N_B}}}{T_{\sigma_{1, N_B}}} \sqrt{N_B} - \frac{\Delta_{H_1, N_B}(s)}{T_{\sigma_{1, N_B}}},
\end{align*}
which results in 
\begin{eqnarray*} 
\sqrt{\xi_B} \frac{T_{\sigma_{0, N_B}}}{T_{\sigma_{1, N_B}}} \sqrt{N_B} - \sqrt{\xi_A} \frac{T_{\sigma_{0, N_A}}}{T_{\sigma_{1, N_A}}} \sqrt{N_A} = \\  \frac{\nu\,!}{s^{\nu}} \left\{ \Phi^{-1}(1 - \alpha)\left[\frac{T_{\sigma_{0, N_B}}}{T_{\sigma_{1, N_B}}} -  \frac{T_{\sigma_{0, N_A}}}{T_{\sigma_{1, N_A}}}\right]  + \frac{\Delta_{H_1, N_A}(s)}{T_{\sigma_{1, N_A}}} \right. \\ \left. - \frac{\Delta_{H_1, N_B}(s)}{T_{\sigma_{1, N_B}}}\right\}.
\end{eqnarray*} 
Re-arranging the terms, we get
\begin{eqnarray} 
\frac{\sqrt{\xi_A}  \sqrt{N_A}} {\sqrt{\xi_B}  \sqrt{N_B}} \!\!\!\!\! &=& \!\!\!\!\! \frac{\frac{T_{\sigma_{0, N_B}}}{T_{\sigma_{1, N_B}}}}{\frac{T_{\sigma_{0, N_A}}}{T_{\sigma_{1, N_A}}}} \left(1 - U\right),
\label{eq:re_are1}
\end{eqnarray}
where 
\begin{eqnarray} 
\nonumber U = \frac{\frac{\nu\,!}{s^{\nu}} \left\{ \Phi^{-1}(1 - \alpha)\left[\frac{T_{\sigma_{0, N_B}}}{T_{\sigma_{1, N_B}}} -  \frac{T_{\sigma_{0, N_A}}}{T_{\sigma_{1, N_A}}}\right] \right\}} {\sqrt{\xi_B} \frac{T_{\sigma_{0, N_B}}}{T_{\sigma_{1, N_B}}} \sqrt{N_B}} + \\
\frac{\frac{\nu\,!}{s^{\nu}} \left\{ \frac{\Delta_{H_1, N_A}(s)}{T_{\sigma_{1, N_A}}} - \frac{\Delta_{H_1, N_B}(s)}{T_{\sigma_{1, N_B}}}\right\}} {\sqrt{\xi_B} \frac{T_{\sigma_{0, N_B}}}{T_{\sigma_{1, N_B}}} \sqrt{N_B}}.
\label{eq:U}
\end{eqnarray} 
Squaring both sides of \eqref{eq:re_are1} and by using the definitions of RE and ARE, we get 
\begin{eqnarray} 
\text{RE}_{A,B} \!\!\!\!\! &=& \!\!\!\!\! \left(\frac{\frac{T_{\sigma^2_{1, N_B}}}{T_{\sigma^2_{0, N_B}}}}{\frac{T_{\sigma^2_{1, N_A}}}{T_{\sigma^2_{0, N_A}}}} \right)  \frac{\text{ARE}_{A,B}}{\left(1 - U\right)^2}.
\label{eq:re_are}
\end{eqnarray}

\section{Remarks}\label{sec:remarks}
Equation \eqref{eq:re_are} is the identical to the formula that connects RE to ARE for constant signals developed in \cite[Equation (8)]{Blum1991}, however there are two main differences. Firstly, the variance terms appearing in \eqref{eq:re_are} are for the random signal detection problem, while those in \cite[Equation (8)]{Blum1991} are for known signal detection. Secondly, the expression for $U$ in \eqref{eq:U} has the unknown parameter $s$, while the expression for $U$ in \cite[Equation (8)]{Blum1991} can be calculated for a fixed value of $s$. For the random signal case, the value of $U$ can only be estimated. 

For a known signal, the first order derivative of the mean of the test statistic under the alternative hypothesis is nonzero for $s = 0$ \cite{Poor1988}. In the limit when $s \sqrt{N_B}$ approaches a constant, for $N \rightarrow \infty$, $s$ approaches $0$ resulting in $U \rightarrow 0$. Thus RE converges to ARE up to a multiplicative constant without the influence of $U$. For $s$ random, this behavior cannot be predicted unless $s$ is estimated accurately. Also, note that, for random signal detection, typically $\nu = 2$ (\cite[Section VI]{Song1990}). It remains to be investigated how this affects the convergence rate of RE to ARE even if $s$ is estimated accurately.

For known $s$, \eqref{eq:re_are} holds for $N \rightarrow \infty$ only when the assumption that $s \rightarrow 0$ is relaxed. When $s$ is random, \eqref{eq:re_are} holds for $N \rightarrow \infty$ provided $\mu_1 \rightarrow 0$. For known $s$, the rate of convergence of RE to ARE is fully determined by the rate of convergence of $U$ to zero, while for $s$ random, the convergence of RE to ARE is largely governed by the accuracy of the estimate of $s$ (and hence, the estimate of $U$). 

For constant $s$, the rate of convergence of RE to ARE is fully determined by the rate of convergence of $U$ to zero. As $N \rightarrow \infty$, we see that $U < 1$, and hence from \eqref{eq:re_are} (for constant $s$) RE approaches ARE from above. If $U < 0$, then RE will be smaller than ARE and approaches ARE from below. For $s$ random, the rate of convergence of RE to ARE is determined not only by the rate of convergence of $U$ to zero, but also by the rate of convergence of the estimator of $s$. For instance, for an estimator $\hat{s}$ based on the sample mean, the impact of the rate of convergence of $\hat{s}$ on the convergence of $U$ to zero would provide insights into the convergence of RE to ARE \cite{Bahadur1967}, \cite{Perng1978}. This paper is only a preliminary report of this convergence analysis; a more comprehensive discussion along with numerical results will be made available in the future.

\bibliographystyle{IEEEtran}
\bibliography{IEEEabrv,enst}
\raggedbottom

\end{document}